\documentclass[12pt]{article}%
\usepackage{graphicx}
\usepackage{amsmath}
\usepackage{amsfonts}
\usepackage{amssymb}%
\newtheorem{theorem}{Theorem}
\newtheorem{acknowledgement}[theorem]{Acknowledgement}

\begin{document}

\title{Provable first-order transitions for liquid crystal and lattice gauge 
models with continuous symmetries. }
\author{Aernout C. D. van Enter\\Institute for Theoretical Physics\\Rijksuniversiteit Groningen \\P.O. Box 800\\9747 AG Groningen \\the Netherlands\\aenter@phys.rug.nl
\and Senya. B. Shlosman\\CPT, CNRS Luminy, Case 907 \\F13288 Marseille Cedex 9 France\\shlosman@cptsu5.univ-mrs.fr}
\maketitle

\textbf{Abstract: We consider various sufficiently nonlinear sigma models for
nematic liquid crystal ordering of $RP^{N-1}$ type and of lattice gauge type 
with continuous symmetries. We rigorously show that they exhibit a first-order 
transition in the temperature. The result holds in dimension 
$\mathbf{2}$
or more for
the $RP^{N-1}$ models and in dimension 
$\mathbf{3}$
or more for the
lattice gauge models. In the two-dimensional case our results clarify and
solve a recent controversy about the possibility of such transitions. For
lattice gauge models our methods provide the first proof of a first-order
transition in a model with a continuous gauge symmetry. 
}

In \cite{EntShl}, it was shown that a class of ferromagnetic $N$-vector models
with a sufficiently nonlinear nearest-neighbour interaction -- meaning that
the nearest neighbour interaction has the shape of a deep and narrow well --
show a first-order transition in temperature. 
An example of a
model which could be
treated is given by the Hamiltonian
\[
H=-J\sum_{{<}i,j{>}\in\mathbb{Z}^{2}}\left(  \frac{1+\cos\left(  \phi_{i}%
-\phi_{j}\right)  }{2}\right)  ^{p},
\]
with 
$p$
large.
These results confirmed earlier numerical work
\cite{DomSchSwe, BloGuoHil}. The main ingredient of the proof was a similarity
between such models and high-$q$ Potts models, which allows one to adapt
proofs for Potts models, first developed in \cite{KotShl}, and based on
\cite{DS}, to prove first-order transitions in the temperature parameter for
these $N$-vector models.

In this paper we extend our analysis to include $RP^{N-1}$ (liquid-crystal)
models (such as were first introduced by Lasher and Lebwohl \cite{LasLeb,Las})
and lattice gauge models. We can then employ techniques for showing the
existence of first-order transitions, such as have been used for Potts
ferromagnets in $d$ at least $2$, as well as for Potts lattice gauge models in
$d$ at least $3$ \cite{KotShl}, to conclude that the corresponding nonlinear
liquid-crystal and lattice gauge models, (with either abelian or non-abelian
symmetries) have 1st order transitions.

The standard 
${N}$-vector models
are either believed
or sometimes rigorously known to have 2nd order transitions in $d=3$ or
higher, a "Kosterlitz-Thouless" transition in $d=2,N=2,$ and no transition for
$d=2$ and higher $N$ \cite{convention}. In the XY-model $(N=2)$ for either
$d=2$ or high $d$ these results are rigorous, for the other models there is a
consensus based on both numerics and heuristic arguments.

In contrast, for the standard versions of the liquid crystal and lattice gauge
models, as well as for very non-linear $\sigma$-models, both numerics and high
temperature series suggested the existence of 1st order transitions, despite
some theoretical analyses originally either suggesting 2nd order transitions,
no transitions at all, or Kosterlitz-Thouless type transitions. For some of
this literature, see e.g. \cite{KunZum2,LasLeb,Las,EspTag,ABLS,Sol,
Pes,Sav,SokSta} and references therein. Moreover, in the limit where $N$
approaches infinity (the spherical limit) 1st order transitions were found, in
dimension $2$ or more \cite{KunZum1,SonTch,SokSta}. This spherical limit
result also holds for our nonlinear interactions in the ferromagnetic case
\cite{CarPel}. Whether such a first-order transition can also occur for finite
$N$ larger than $3$ in $d=2$, or whether it might be an artefact of the
spherical limit has for a long time been a matter of controversy (see for
example \cite{SokSta, SonTch}). In fact, Sokal and Starinets described the
existence of such a first-order transition as a \textquotedblleft
pathology\textquotedblright. Our result finally settles this question:
first-order phase transitions for models with continuous symmetry in $d=2$ can
occur, despite the conjecture to the contrary of \cite{SokSta}. Our results
are essentially in agreement with the analysis of \cite{SonTch}. In contrast
to what was suggested in most earlier analyses, the symmetry or the
low-temperature properties of the model do not play a role of any great
importance, and in fact for our nonlinear choice of interaction the
spin-dimensionality $N$ does not need to be large and can be as small as $2$.
The lattice dependence of the phenomenon found in \cite{SonTch} seems somewhat
of an artifact, however, which disappears if one varies the nonlinearity 
parameter.

The fact that our proofs are insensitive to the nature of the phases between
which the transition takes place implies that, similarly to the ferromagnetic
cases, one might have in the liquid-crystal models a transition between a
disordered high-temperature phase and either a nematically-ordered, a
Kosterlitz-Touless or a disordered phase at low temperatures. Similarly one
might find a transition either between a confining and a nonconfining --
Coulomb-like -- phase or between two confining phases in the lattice gauge
models. Which one occurs in a particular case should depend on dimension
and{/}or symmetry of the system.

In particular, we emphasize that our proof is also insensitive as to whether
the symmetry group of the lattice gauge model is abelian -- in which case it
is expected that in $4$ dimensions a transition between a confined and a
Coulomb-like phase occurs \cite{Guth, FS1} --, or nonabelian, in which case
both states are expected to confining, (this is also expected in general in
$d=3$). For a heterodox discussion on the difference between what is to be
expected in abelian and nonabelian models, including some history of this
problem cf \cite{PatSei}.

We consider a lattice $\mathbb{Z}^{d}$, and either spin models in which the
random variables live on the sites, or lattice gauge models where the
variables live on the bonds (or links) of the lattice. The parameters of our
models are the spatial dimension $d$, the spin-dimension $N$, and the
nonlinearity parameter $p$. The \textquotedblleft standard\textquotedblright%
\ versions of the models are obtained by taking $p=1$.

For ferromagnetic models the variables are $N$-component unit vectors living
on the $N$-sphere. We will discuss the argument in the $2$-component case, in
the general case the proof is essentially the same. The ferromagnetic models
treated in \cite{EntShl} were given by
\begin{equation}
H=-J\sum_{{<}i,j{>}\in\mathbb{Z}^{2}}\left(  \frac{1+\cos\left(  \phi_{i}%
-\phi_{j}\right)  }{2}\right)  ^{p}. \label{21}%
\end{equation}

For liquid crystal $RP^{N-1}$ models either one considers variables -- usually
denoted $n_{i}$ -- which live on the projective manifold obtained by
identifying a point on the $N$-sphere with its reflection through the origin,
or equivalently one can consider ordinary spins on the $N$-sphere, and divide
out this \textquotedblleft local gauge symmetry\textquotedblright\ afterwards.
The last approach is the route we will pursue, as it allows us to literally
transcribe the proof of \cite{EntShl}. Thus we consider Hamiltonians of the
form
\begin{equation}
H=-J\sum_{{<}i,j{>}\in\mathbb{Z}^{2}}\left(  \frac{1+\cos^{2}\left(  \phi
_{i}-\phi_{j}\right)  }{2}\right)  ^{p}.
\end{equation}
In \cite{EntShl} we call a bond \textquotedblleft ordered\textquotedblright%
\ if the angle $\theta$ between two neighboring sites is small enough. Here we
call it ordered if the angle $\theta\mathrm{mod}\pi$ is small enough. Then the
argument of \cite{EntShl}, based on \cite{Shl} and \cite{KotShl} goes through
without any changes. There is a first-order phase transition for $p$ chosen
large enough (in general the values of $p$ for which the proof works depend on
$N$ and $d$) between a high-temperature regime in which most nearest neighbor
bonds are disordered, and a low-temperature regime in which most nearest
neighbor bonds are ordered. This holds for each dimension at least $2$, and
whereas the Mermin-Wagner theorem excludes nematic long-range order in $d=2$
\cite{MerWag}, in $d=3$ and higher long-range order will occur \cite{AngZag}.
Between the ordered and the disordered phase(s) free energy contours occur,
whose probabilities are estimated to be uniformly small via a contour estimate
valid over a whole temperature interval. In the contour estimate use is made
of the Reflection Positivity \cite{RP} of the model. We remark that at low
temperatures percolation of ordered bonds holds \cite{Geo}; it follows from
our results that the associated percolation transition is first order.

For lattice gauge models the variables are elements of a unitary
representation of a compact continuous gauge group, e.g. $U(1),\ SU(n)$, or
sums thereof \cite{Smi}. Here we present the argument in the simplest case of
a $U(1)$-invariant interaction in $3$ dimensions.%

\begin{equation}
H=-J\sum_{plaquettes\ P\in\mathbb{Z}^{3}}L\left(  U_{P}\right)  ,
\end{equation}
with $L(U_{P})=\left(  \frac{1+\cos\left(  \phi_{e_{1}}+\phi_{e_{2}}%
-\phi_{e_{3}}-\phi_{e_{4}}\right)  }{2}\right)  ^{p}.$ Here the $e_{i}$ denote
the 4 edges making up the plaquette $P$.

The effect of choosing the nonlinearity parameter $p$ high is that the
potential, although it still has quadratic minima, becomes much steeper and
narrower. In this way one constructs in a certain sense a ``free energy
barrier'' between ordered and disordered phases.

The lattice gauge model proof becomes similar to the arguments from
\cite{KotShl}. When the product over the link variables is sufficiently close
to unity, we'll call the plaquette \textquotedblleft ordered\textquotedblright%
, \textquotedblleft disordered\textquotedblright\ otherwise. This distinction
corresponds to unfrustrated and frustrated plaquettes in the Potts case. We
will sketch the argument for the toy model where the potential $L(U)$ is
chosen to be $1$ if the sum $\phi_{P}$ of the oriented angles along the
plaquette $P$ is between $-\frac{\mathbb{\varepsilon}}{2}$ and $+\frac
{\mathbb{\varepsilon}}{2}$ and zero otherwise. The generalization to the
non-square well potentials can then be done in the same way as in
\cite{EntShl}. The correspondence, as in \cite{EntShl}, is that 
${\mathbb{\varepsilon}}$ is of order $O(\frac {1}{\sqrt p})$.

Our strategy is to find bounds for free energy contours between ordered
phases, in which one has mainly cubes with 6 ordered plaquettes, and
disordered phases, in which most cubes have 6 disordered plaquettes. We need
thus to estimate the weights of contours consisting of cubes which are neither
ordered nor disordered. The number of possibilities for such cubes includes
the 7 possibilities given in \cite{KotShl}, except that now we have the
additional 8th possibility of having cubes with one disordered plaquette on
one side, and five ordered ones.

For the partition function $Z_{L}$ on a cube $B_{L}$ of size $L^{3}$ we use
the 
(quite rough)
lower bound
\begin{equation}
Z_{L}\geq max(Z_{L}^{d},Z_{L}^{o}),
\end{equation}
where $Z_{L}^{d}$ (resp., $Z_{L}^{o}$) is part of the partition function,
calculated over all configurations which have all plaquettes disordered
(resp., 
mostly
ordered). 
For the disordered partition
function 
$Z_{L}^{d}$
we obtain the lower bound 
$\left(1-4\varepsilon\right)  ^{3L^{3}}$
%
(we take a normalized reference
measure, giving a weight 
$1$
to each link).

For the ordered partition function $Z_{L}^{o}$ we proceed as follows: we first
choose a set of bonds $T_{L}$ in $B_{L},$ which is a tree, passing through
every site. For example, we can put into $T_{L}$ all vertical bonds --
$z$-bonds -- except these connecting sites with $z$-coordinates $0$ and $1,$
plus all $y$-bonds in the plane $z=0,$ except these connecting the sites with
$y$-coordinates $0$ and $1,$ plus all $x$-bonds of the line $y=z=0,$ except
the one between the sites $\left(  0,0,0\right)  $ and $\left(  1,0,0\right)
.$ The site $\left(  0,0,0\right)  $ can be taken as a root of $T_{L}.$ Note
that the number of bonds in $T_{L}$ is $L^{3}-1.$ Therefore it is not
surprising (and easy to see) that for every edge configuration $\mathbf{\phi
}=\left\{  \phi_{b},b\in T_{L}\right\}  $ there exists a unique site
configuration $\mathbf{\psi}=\mathbf{\psi}_{\mathbf{\phi}}=\left\{  \psi
_{x},x\in B_{L}\right\}  ,$ such that the following holds:

\begin{enumerate}
\item Let $\mathfrak{g}^{\psi}$ denote the gauge transformation, defined by
the configuration $\psi.$ Then $\left(  \mathfrak{g}^{\psi}\circ\mathbf{\phi
}\right)  \Bigm|_{b}=1$ for every bond $b\in T_{L};$

\item $\psi_{\left(  0,0,0\right)  }=1.$
\end{enumerate}

\noindent For every family of bonds $S\subset B_{L}$ let us define a bigger
family $C\left(  S\right)  ,$ by the rules:

\begin{enumerate}
\item $S\subset C\left(  S\right)  ,$

\item for every four bonds $\left\{  b_{1},...,b_{4}\right\}  ,$ making a
plaquette, such that three of them are in $S,$ we have $\left\{
b_{1},...,b_{4}\right\}  $ $\subset C\left(  S\right)  .$
\end{enumerate}

\noindent Then we can consider also the sets $C^{2}\left(  S\right)  =C\left(
C\left(  S\right)  \right)  ,$ $C^{3}\left(  S\right)  ,$ and so on. Define
$\mathfrak{C}\left(  S\right)  =\cup_{k}C^{k}\left(  S\right)  .$ Note that
the number of plaquettes in $\mathfrak{C}\left(  T_{L}\right)  $ is
$3L^{3}-O\left(  L^{2}\right)  .$ We claim now that for every configuration
$\mathbf{\phi}_{T_{L}}=\left\{  \phi_{b},b\in T_{L}\right\}  $ one can specify
(in a continuous way) a collection of arcs

\noindent$\left\{  I_{b}=I_{b}\left(  \mathbf{\phi}_{T_{L}}\right)  \subset
S^{1},b\in\mathfrak{C}\left(  T_{L}\right)  ~\backslash~T_{L},\left\vert
I_{b}\right\vert =\frac{\varepsilon}{4}\right\}  ,$ such that for every
configuration $\mathbf{\phi}$ on $B_{L},$ which coincides with $\mathbf{\phi
}_{T_{L}}$ on $T_{L},$ and for which the values $\phi_{b}$ on the bonds
$b\in\mathfrak{C}\left(  T_{L}\right)  ~\backslash~T_{L}$ belong to the above
segments $I_{b},$ all the plaquettes that fall into $\mathfrak{C}\left(
T_{L}\right)  $ are non-frustrated. That would imply that
\[
Z_{L}^{o}\geq\left(  \frac{\varepsilon}{4}\right)  ^{2L^{3}}\exp\left\{
3J\left(  L^{3}-O\left(  L^{2}\right)  \right)  \right\}
\]
by Fubini's theorem. To see the validity of our claim, consider first the
case when the configuration 
$\mathbf{\phi}_{T_{L}}\equiv\mathbf{1}\in S^{1}$ (here
$1$ is the neutral element). Then the choice of the segments $I_{b}$ is easy:
$I_{b}\left(  \mathbf{1}\right)  =\left[  -\frac{\varepsilon}{8}%
,\frac{\varepsilon}{8}\right]  $ for every $b\in\mathfrak{C}\left(
T_{L}\right)  ~\backslash~T_{L}.$ For a general $\mathbf{\phi}_{T_{L}}$ let us
take the corresponding gauge transformation $\mathfrak{g}^{\mathbf{\phi
}_{T_{L}}}$ (which is the identity for
$\mathbf{\phi}_{T_{L}}\equiv\mathbf{1}$),
and we define our segments by
\[
I_{b}\left(  \mathbf{\phi}_{T_{L}}\right)  =\left(  \mathfrak{g}%
^{\mathbf{\phi}_{T_{L}}}\right)  ^{-1}I_{b}\left(  \mathbf{1}\right)  .
\]

This provides a lower bound%

\begin{equation}
Z_{L}\geq\max\left[  \left(  1-4\varepsilon\right)  ^{3L^{3}},\left(
\frac{\varepsilon}{4}\right)  ^{2L^{3}}\exp\left\{  3J\left(  L^{3}-O\left(
L^{2}\right)  \right)  \right\}  \right]  .
\end{equation}

This bound on the partition function as the maximum of the ordered and
disordered term is similar to the argument in \cite{EntShl}. It plays the same
role as the bound in terms of a fixed energy partition function given in
\cite{KotShl}.

To obtain our contour estimates, by Reflection Positivity we need to estimate
the partition functions of configurations constrained to have a
\textquotedblleft universal contour\textquotedblright. The estimates of the 7
types of universal contours mentioned in \cite{KotShl} are of a similar form
as in that paper with the number of Potts states $q$ up to some constants
replaced by $ \frac{1}{\varepsilon}$. The universal contour due to the new
case of
cubes with one disordered plaquette consists of configurations in which the
horizontal plaquettes in every other plane are disordered, and all the
other ones
are ordered. These configurations have a similar entropy contribution to
the partition function as the ordered configurations, but the energy per
cube
is $\frac{5}{6}$ of that of a cube in the fully ordered situation. For
$\mathbb{\varepsilon}$ small enough (which corresponds to $p$ large enough)
also such a contour is suppressed exponentially in the volume. The
combinatorial factor in the contour estimate changes by some finite constant,
which choosing $\mathbb{\varepsilon}$ small enough takes care of.

To summarize we have obtained the following results:

\begin{theorem}
For any nonlinear $RP^{N-1}$ model in dimension $2$ or more and $p$ high
enough, there is a first order transition, that is, there exists a
temperature at
which the free energy is not differentiable in the temperature parameter.
\end{theorem}

\begin{theorem}
For lattice gauge models with plaquette action $\left(  \frac{1+L(U_{P}%
)}{2}\right)  ^{p}$, (where $L(U_{P})=Tr(U_{P}+U_{P}^{\ast})$) in dimension
$3$ and more and $p$ high enough, there is a first order transition, that is
there exists a temperature at which the free energy is not differentiable in
the temperature parameter.
\end{theorem}

Here $U_{P}^{\ast}$ denotes the adjoint operator of $U_{P}$.

This seems to be the first case in which a first order transition for a
lattice gauge model with a continuous gauge symmetry group is rigorously
obtained. Whereas the example of the Potts lattice gauge model in $d=3$ or
higher is between a confining and a nonconfining phase \cite{KotLaaMesRui,
LaaMesRui}, in our theorem this is to be expected in $d=4$, with $U(1)$
symmetry only. For $d=3$ and also for $SU(n)$ in $d=4$ we conjecture that
confined phases exist on both sides of the phase transition.

Our proof only gives results for very high $p$. We will discuss some further
aspects of what may actually be the $p$-values for which to expect first-order
transitions, and what one might hope to prove. The recent work of Biskup and
Chayes \cite{BisCha} shows that if a reflection positive model has a phase
transition in mean field theory, then also at sufficiently high dimension a
first-order transition occurs. They include in their discussion the $RP^{N-1}$
model for $N=3$, for which even for $p=1$, a first-order transition is
derived. The mean field analysis of \cite{BloGuo} indicates that a similar
result for the ferromagnetic case holds if $p=3$. For lattice gauge models
also the standard actions lead to first order transitions in mean-field theory
\cite{Zin}, section 34.4, which indicates a first order transition in
sufficiently high dimension.

If one believes that here the spherical ($N$ to infinity) limit is not
singular (which has been a matter of controversy itself), then for the square
lattice, $N$ large and $p$ larger than $6$ the ferromagnet might have a first
order transition, while for the $RP^{N-1}$ case on the square or triangular
lattice even for $p=1$ a first order transition occurs, while for the
hexagonal lattice one presumably needs a higher value of $p$ \cite{SokSta,
SonTch}.

As mentioned before, numerical work suggests that in the standard ($p=1$)
Lebwohl-Lasher model with $N=3$ as well as in the $U(1)$-lattice-gauge
model a
first order transition should occur; however, this seems far away from any
provable result.

\begin{acknowledgement}
We thank in particular E.~Domany and A.~Schwimmer who suggested to us to
consider lattice gauge models, and also D.~v.d. Marel, A.~Messager,
K.~Netocn\'{y} and A.~Sokal for stimulating discussions and{/}or correspondence.
\end{acknowledgement}

\end{document}